\documentclass[sigconf]{acmart}
\renewcommand\footnotetextcopyrightpermission[1]{} 
\usepackage{fancyhdr}
\fancypagestyle{plain}{%
\fancyhf{} 
\fancyfoot[C]{\large\thepage} 

}
\usepackage{booktabs} 
\setcopyright{rightsretained}

\begin{document}

\title{SHARVOT: secret SHARe-based VOTing on the blockchain}

\author{Silvia Bartolucci}
\affiliation{%
  \institution{nChain}
  \city{London} 
  \state{UK} 
}
\email{silvia@ncrypt.com}
\author{Pauline Bernat}
\affiliation{%
  \institution{nChain}
  \city{London} 
  \country{UK}}
\email{pauline@ncrypt.com}

\author{Daniel Joseph}
\affiliation{%
  \institution{nChain}
  \city{London} 
  \state{UK} 
}
\email{daniel@ncrypt.com}

\begin{abstract}
Recently, there has been a growing interest in using online technologies to design protocols for secure electronic voting. The main challenges include vote privacy and anonymity, ballot irrevocability and transparency throughout the vote counting process. The introduction of the blockchain as a basis for cryptocurrency protocols, provides for the exploitation of the immutability and transparency properties of these distributed ledgers. 

In this paper, we discuss possible uses of the blockchain technology to implement a secure and fair voting system. In particular, we introduce a secret share-based voting system on the blockchain, the so-called SHARVOT protocol\footnote{SHARVOT's technologies are the subject of the following UK patent applications:\\
\textbf{1703562.7} (06/03/2017). \textbf{1713800.9} (29/8/2017).}. Our solution uses Shamir's Secret Sharing to enable on-chain, \textit{i.e.} within the transactions script, votes submission and winning candidate determination. The protocol is also using a shuffling technique, \textit{Circle Shuffle}, to de-link voters from their submissions. 
\end{abstract}


\maketitle

\section{Introduction} \label{sec:intro}
A blockchain is a public ledger, managed by a peer-to-peer network, offering security, data immutability and transparency to its users ~\cite{Antonopoulos}.
The advent of blockchain technology via the cryptocurrency Bitcoin \cite{Satoshi} has paved the way for the exploitation of the appealing ledger's properties for the  most diverse applications. From healthcare to finance, blockchain-based solutions are created, using digital identification and smart contracts.

In the sector of electronic voting and rating, numerous voting platforms and startups focusing on building election systems on the blockchain have emerged and attracted high attention and funding \footnote{\url{http://www.europarl.europa.eu/RegData/etudes/ATAG/2016/581918/EPRS_ATA(2016)581918_EN.pdf} Retrieved on Feb 4 2018.}. Indeed,
features such as \textit{irrevocability} and \textit{transparency} of information stored on chain make the  technology ideally suited for the development of secure voting systems.

Decentralisation is one of the biggest challenges. Usually existing solutions rely on a central authority to validate and correctly count votes or feedback. The central authority may use different cryptographic primitives, such as blind signatures and homomorphic encryption, to make the users' submissions anonymous and verifiable \cite{review1}. In Chaum's protocol \cite{mixnet}, participants send their votes to mixing authorities that reshuffle and encrypt the votes before broadcasting them, in order to disguise the link between votes and voters. Among the existing decentralised cryptographic solutions, \textit{Secure Multi-Party Computations} protocols enable a set of users to compute a function of their joint private inputs (which they want to keep secret) without the need of a trusted third party, allowing the parties to jointly compute the average of their ratings privately and securely \cite{MPC}.

Focusing on how to preserve votes secrecy while guaranteeing integrity of the process is also fundamental. 
In \cite{ZKPvote}, participants submit binary votes $v_i \in \{0,1\}$ masked by adding a conveniently created random variable together with a proof that the vote is either $0$ or $1$. Despite the votes being concealed, one can verify the vote format and prevent double submissions at the price of using computationally expensive zero-knowledge proofs \cite{RSA}.

In this paper, we present solutions using the blockchain technology to announce and build an election and determine a winning candidate. In particular, we focus on how to broadcast the votes within the transactions script and correctly count them while (i) protecting voters privacy and anonymity (without the need of zero-knowledge proofs), (ii) allowing only eligible users to cast their preference, (iii) preventing attacks aimed at invalidating the ballot.

The paper is organised as follows. In Sec. ~\ref{sec:issues}, we describe the main challenges faced when building a secure and fair voting protocol. In Sec.~\ref{sec:SHARVOT}, we present the \textit{SHARVOT} protocol, a blockchain-based solution using the technology presented in Sec.~\ref{sec:issues}.
Sec. ~\ref{sec:end},~\ref{sec:end2} are devoted to discussion and conclusive remarks.

\section{E-Voting: building blocks} \label{sec:issues}
In this section, we describe how to exploit the features of the ledger and blockchain transactions to cast and correctly count ballots. We discuss possible ways to store the ballot in-script and the vote format respectively in Sec. \ref{sec:datast} and \ref{sec:vct}. In Sec. \ref{sec:elig} and Sec. \ref{sec:shuffle} we describe how to prevent submission from non-eligible voters and how to protect vote's secrecy.

\subsection{Vote recording and data storage}\label{sec:datast}
Data storage on the blockchain is controversial. A conspicuous fraction of developers believe that data may overload the full nodes by increasing (i) disk storage costs and (ii) memory of the \textit{unspendable transaction output} (UTXO) set. Indeed, some applications create pseudo-payments to use the recipient's Bitcoin address as a $20$-bytes field, hence creating unspendable UTXO. According to the Bitcoin protocol, data can instead be stored in script using the opcode $OP\_RETURN <data>$, which allows for the storage of $80$ bytes of non-payments related data \cite{OP-ret,Antonopoulos}.

An alternative solution exploits the pay-to-script-hash (P2SH) script\footnote{P2SH scripts allow a user to lock UTXOs in a transaction in such a way that the recipient must provide data to complete the script, therefore unlocking the UTXOs sent to the Bitcoin address derived from the P2SH script.} of a \textit{m-of-n } multisignature transaction. According to these scripts, $n$ public keys $P_i,~i = 1,\ldots,n$ are recorded, and in order to redeem the script, at least $m \leq n$ signatures $S_j,~j = 1,\ldots,m$, (corresponding to any subset of $n$ public keys) must be provided \cite{Antonopoulos}.
The \textit{m-of-n} multisignature script is of the format:
\begin{equation}
OP\_0 \,S_1\dots S_m <m\, P_1\dots P_n\, n\, OP\_CHECKMULTISIG> \ .
\label{eq:script1}
\end{equation}
The space devoted to public keys can be used to store metadata. Indeed, a multisignature address can be of the form $\textit{q-of-n}$ with $q \geq 1$, where the output can be spent by providing the signatures associated to the $q$ public keys while the remaining $n-q$ public key slots are used to encode data of a maximum size of $64$ bytes each. This eliminates the need for an additional unspendable UTXO and allows anyone to access and verify the metadata stored in the multisignature script. 
Consider for instance a \textit{1-of-3} multisignature script, where two of the three data elements reserved for public keys, are used to store metadata instead. The script takes the following format:
\begin{equation}
OP\_0 \, S_1\,  <OP\_1 \, m_1 \, m_2 \, P_1 \, OP\_3 \, OP\_CHECKMULTISIG>\ ,
\label{eq:script2}
\end{equation}
where $m_1,m_2$ are the metadata stored in the script.
In voting protocols, the metadata elements could correspond to the set of encrypted/shuffled votes $\{v_i\}, i=1,\ldots,m$, of $m$ users.
Note the caveat that, for the current version of the Bitcoin protocol, the maximum number of public keys allowed in a multisignature output is 15, hence limiting the maximum number of votes that can be stored per multisignature script.

\begin{figure*}[ht]
  \centering
  \includegraphics[width=0.7\textwidth]{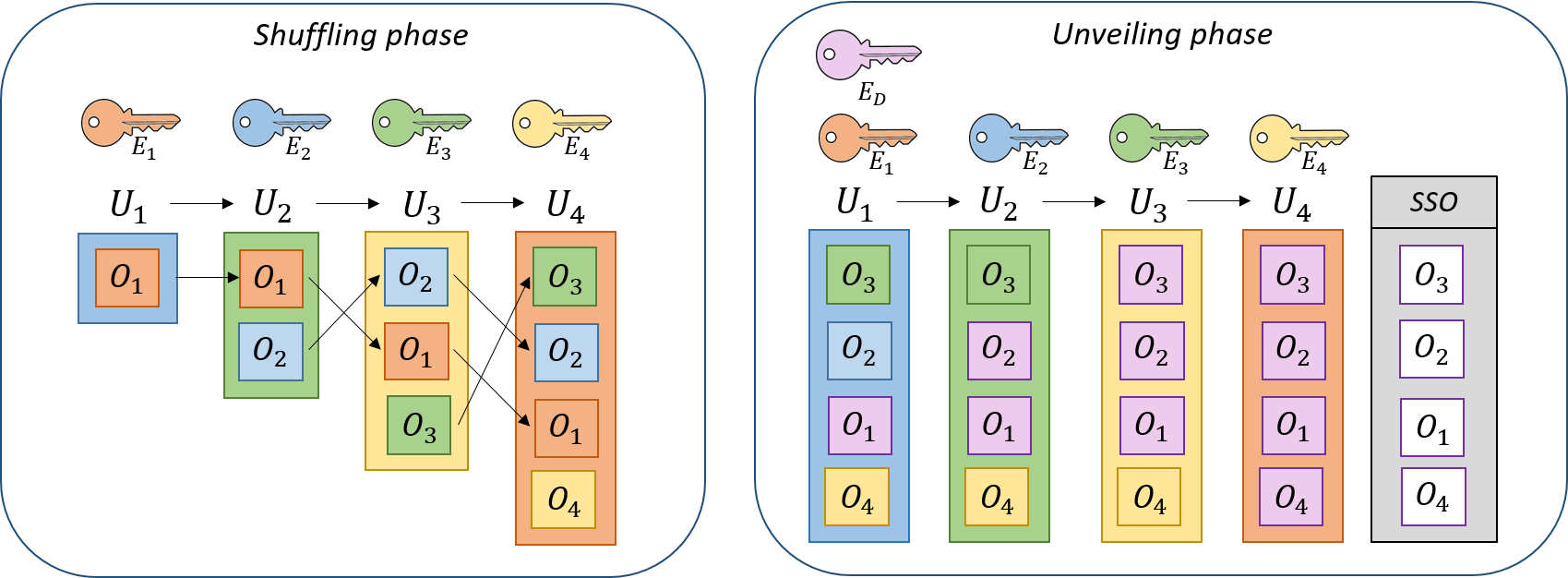}
  \caption{$\textit{Circle Shuffle}$ technique: (left) Shuffling of the encrypted output addresses. (right) Unveiling of the shuffled set of output addresses. $O_i$ and $E_i, i=1,\ldots,4$, represent, respectively, the output and the public key used to encrypt it of the \textit{i}-th participant.}
\label{fig:cs}
\end{figure*}
\subsection{Vote commitment format}\label{sec:vct}
In general, a vote can be submitted in different formats: as a string to indicate the name of a candidate, as a binary variable, an integer, etc. In this paper, we express the vote as a $64$-bytes key, denoted by $k_{i,j}$, where the index $i$ relates to the candidate selected by the user while $j$ identifies a particular voter. We describe how to construct these keys in the next section.

A voter $U_i$ wishing to vote for candidate $C$ submits their key share $k_{C,i}$. The key can also be concatenated with an identification value for candidate $C$ indicated with \texttt{IdC}.
As an example, a vote submission may take the following form:
$v_i=k_{(C,i)} || \texttt{IdC}$
where $||$ represents strings concatenation.
The vote $v_i$ can also be encrypted with the public key $Pk_C$ of the candidate before being embedded in the script, \textit{e.g.} $\hat{v_i}=Enc_C (v_i)$.
In the event that the candidate $C$ attempts to decrypt the vote, the component \texttt{IdC} serves as an indicator to the candidate that the decryption was successful. 

\subsection{Voting using Shamir$\textquotesingle$s Secret Sharing} \label{subsec:shamir}
$\it{Secret}$ $\it{Sharing}$ schemes are examples of $t$-of-$n$ threshold cryptosystems, whereby a secret $k$ is divided among $n$ participants, such that at least $t+1$ participants are required to collaborate to reconstruct $k$. The knowledge of any $t$ pieces of the secret $k$ leaves the latter undetermined.

Shamir$\textquotesingle$s secret sharing \cite{Shamir} is based on $\it{Lagrange}$ $\it{Polynomial}$ $\it{Interpolation}$ and the secret $k$ is assumed to be an element of a finite field $F$ of size $p$ ($p$ is a prime number). The scheme comprises a dealer (dealer-less versions also exist) and a set of $n$ participants $U_1, U_2, \dots, U_n$.

In the presence of a dealer, the latter chooses an arbitrary $t$-degree polynomial $f(x)=\sum_{j=0}^{t}a_j x^j$ and defines the secret as $k=f(0)=a_0$. The dealer then computes $n$ points $p_i=(x_i,f(x_i)), i=1,\ldots,n$, and distributes the point $p_i$ to the participant $U_i$, i.e. each participant receives their share $k_i= f(x_i)$ of the secret. Using the $\textit{Lagrange Polynomial Interpolation}$, if $t+1$ out of the $n$ participants collaborate, they can reconstruct any point on $f(x)$ with their shares of the key $k$. 

In Sec. \ref{sec:SHARVOT} we show how key shares can be used to express a preference and determine the winner in an election process.

\subsection{Eligibility}\label{sec:elig}
E-voting systems are often targeted by \textit{Sybil attacks}, where a single user attempts to gain access to multiple identities and modify the outcome of the election.
In particular, on the blockchain users can create multiple pseudonyms, \textit{i.e.} public keys. To prevent this kind of attacks, one could (i) introduce a fee per vote, which would discourage an attacker from disrupting the process (the attack might be less cost-effective); (ii) introduce an identification system to allow only registered voters to submit their ballot.

In the second case, a personal secret key $k_i, i=1,\dots,n$, may be given to a set of $n$ endorsed users by a trusted authority, \textit{e.g.} a company announcing an internal election, a website etc. Endorsed voters use their private key to encrypt their vote. The list of keys is kept private by the verifier who later validates the submission: only allowed users will have their vote correctly decrypted and counted.

A possible submission of vote $v_i$ is of the form $ v_i \oplus k_i$, where $\oplus$ represents the XOR of the two strings. Since $v_i\oplus k_i \oplus k_i= v_i$, the verifier decrypts the vote by XOR-ing the string with the key of the users $k_i$. The decryption would only be available to users whose key is known to the verifier.

 Alternatively, the verifier can create a pair of public/private key per user, $(Pk_i, Sk_i)$, and communicate to the eligible user the public key only. The user can encrypt their ballot using $Pk_i$. The vote submission, of the form $Enc_{Pk_i}(v_i)$, is sent to the verifier who decrypts it using the private key ${Sk_i}$, $Dec_{Sk_i}(Enc_{Pk_i}(v_i))=v_i$ \footnote{The functions $Dec_{Sk_i},Enc_{Pk_i}$ vary depending on the cryptosystem used. For instance, in the RSA \cite{RSA}, 
 given a public key $(n,e)=k$, the private key $d$ and the plaintext to encrypt $x$, $Enc_{k}(x)=x^e {\rm{mod}}\ n$ and $Dec_{d}(Enc_{k}(x))= (x^e {\rm{mod}} \ n)^d$.}.

\subsection{De-linking users and submissions}\label{sec:shuffle}
Research on de-anonymisation techniques shows that, despite the use of pseudonyms in cryptocurrencies networks such as Bitcoin, users may be exposed to attacks aimed at uncovering real identities ~\cite{attacks, mix}. To address this vulnerability, a variety of $\textit{coin mixing}$ solutions have been developed to enhance the untraceability of coins flows and the unlinkability of transactions with the coins owners ~\cite{Antonopoulos}. These solutions typically disguise the links between users' addresses by mixing the individual input and output of users in one single transaction, so-called $\it{CoinJoin}$ transaction ~\cite{coinj}. 

We developed a new decentralised solution, $\textit{Circle Shuffle}$, for randomizing the position of the output addresses that compose a $\textit{CoinJoin}$ transaction. This protocol is designed in such a way that no participant signing the transaction knows which of the output addresses corresponds to a specific Bitcoin input address - therefore addressing a limitation present in the $\textit{CoinShuffle}$ protocol, where at least one party in the construction of the transaction has knowledge of the input and output address of another party ~\cite{coinsh}. In the context of voting protocols, shuffling techniques are useful to de-link users from their votes, as we describe in Sec.~\ref{sec:SHARVOT}.

\subsubsection{Initialisation} 
A random sequence $S$ of $n$ participants \\ $U_1, U_2, \dots, U_n$ is decided. Each participant $U_i$ has a corresponding public/private key pair: ($E_i, k_i$) where $E_i = k_i \times G$ ~($G$ is the generator of the Elliptic Curve chosen in the protocol \cite{RSA}). The public keys $E_1, E_2, \dots, E_n$ are made available to all participants. Additionally, the first participant publishes the public key of an ephemeral pair: ($E_D, k_D$).

\subsubsection{Shuffling Phase}
The shuffling phase begins when the first participant, $U_1$, encrypts their output address $O_1$ with their public key $E_1$. The encrypted output address then comprises the `Set of Shuffled Outputs' ($\it{SSO}$). 

$U_1$ encrypts the $\it{SSO}$ with $E_2$, the public key of the next participant in $S$, $U_2$,  and forwards the encrypted $\it{SSO}$ to $U_2$. $U_2$ then decrypts the $\it{SSO}$ using $k_2$, encrypts their output address $O_2$ using $E_2$, and adds the newly encrypted address to the $\it{SSO}$. $U_2$ shuffles the order of the encrypted output addresses within the $\it{SSO}$, encrypts the shuffled $\it{SSO}$ with $E_3$, and then forwards the encrypted $\it{SSO}$ to the third participant $U_3$. This process continues until it reaches $U_n$. This phase is illustrated in Fig.\ref{fig:cs} (left) for a set of $n=4$ participants.

Once the encrypted output address of the last participant is added to the $\it{SSO}$, $U_n$ performs a final shuffle, encrypts the $\it{SSO}$ with $E_1$, and sends the $\it{SSO}$ back to the first participant, $U_1$.
\subsubsection{Unveiling Phase} 
At this point, the $\it{SSO}$ is once again sent to each of the $n$ participants. The ephemeral public key $E_D$, generated by the first participant $U_1$, is used to ensure that no participant gains information about the others during the protocol directly. 

At the end of the first encryption-shuffle loop, the first participant $U_1$ is in possession of the set of shuffled outputs. $U_1$ decrypts the $\it{SSO}$, searches for their encrypted output address in the $\it{SSO}$ and, when found, decrypts the address using the associated private key $k_1$. 
The first participant then encrypts $O_1$ with $E_D$. $U_1$ then encrypts the new $\it{SSO}$ with the public key of the second participant, $P_2$, and forwards the encrypted set to $U_2$. The second participant in turns decrypts the $\it{SSO}$ using $k_2$, finds their output address, decrypts it using $k_2$ and re-encrypts it with $E_D$. $U_2$ then encrypts the new $\it{SSO}$ with the public key $P_3$ and forwards the encrypted set to $U_3$. The process continues until each participant has found their encrypted output address in the $\it{SSO}$ and replaced it with the corresponding decrypted (and encrypted with $E_D$) value, see Fig. \ref{fig:cs} ~(right).

The last participant encrypts the $\it{SSO}$ with $E_1$ (or $E_D$) and sends the encrypted $\it{SSO}$ to the first participant $U_1$.
The first participant decrypts the $\it{SSO}$, and each encrypted address contained within it, using the ephemeral private key $k_D$. The permutation of the output addresses in the $\it{SSO}$ at that point represents the final order in which the outputs are included in a $\it{CoinJoin}$ transaction. 

\begin{figure*}[ht]
  \centering
  \includegraphics[width=0.95\textwidth]{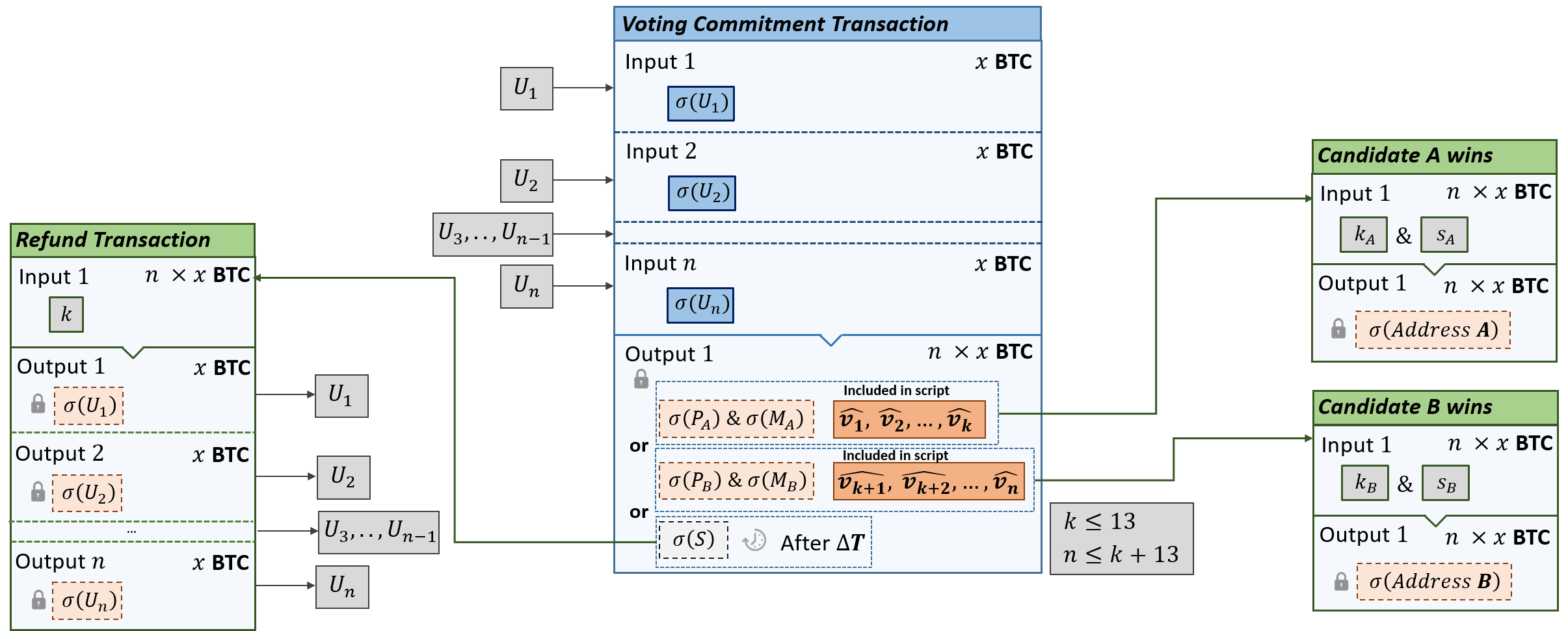}
\caption{Transactions created in the $\it{SHARVOT}$ protocol: the \textit{Voting Commitment Transaction} casting the votes and spending the voters' fees, the transactions owned by each candidate, and the $\it{Refund Transaction}$ returning the fees to the voters in case no candidate wins.}
\label{fig:tssv}
\end{figure*}
\section{The SHARVOT protocol} \label{sec:SHARVOT}
Consider $n$ voters $U_1, \dots, U_n$, wishing to express their preference between several candidates. For simplicity, we will assume a choice between two candidates, $A$ and $B$. The \textit{SHARVOT} protocol consists in a voting solution utilising the Bitcoin blockchain; the protocol is based on the Shamir's secret sharing scheme and a shuffling technique, the \textit{Circle Shuffle}. 

Private key shares (based on the scheme presented in Sec. $\ref{subsec:shamir}$) are assigned to the voters who commit fees in a $n$ inputs - $1$ output transaction that stores, and therefore permanently records the votes. Each input corresponds to the fees paid by a voter. The winning candidate must receive both a majority of votes and a number of votes past a specified threshold to collect the fees.

The $\it{SHARVOT}$ protocol is designed with a failsafe such that any Bitcoins committed by the voters is recoverable if no candidate collects enough key shares to spend the UTXO in the transaction signed by the voters. A dealer-based key share distribution scheme is expected to be the most popular implementation of the $\it{SHARVOT}$ protocol, as having a dealer assume the role of an authority control over the list of eligible voters.
\paragraph{Keys Generation}
A public/private key pair is assigned by the dealer to each of the two candidates, A and B, respectively $P_A=k_A \times G$ and $P_B=k_B \times G$. The candidates also possess a public/private key pair each, $M_A=s_A \times G$ and $M_B=s_B \times G$ respectively, and publish their public key. The dealer also publishes the public keys $P_A $ and $P_B$ while keeping $k_A$ and $k_B$ private. For each secret, the dealer computes $n$ key shares, {$k_{A,i}, i = 1,\ldots,n$,} and {$k_{B,i}, i =1,\ldots,n $}, and distributes a pair of key shares ($k_{A,i}$, $k_{B,i}$) to each voter $U_i$. The key shares are obtained using the $t$-of-$n$ threshold scheme presented in Sec. $\ref{subsec:shamir}$. As a result, the $k_A$ and $k_B$ values can only be determined when voters submit a sufficient number of votes $t+1$ for a particular candidate.
\paragraph{Votes Submission}Each voter $U_i$ encrypts their ballot for the candidate of their choice, as explained in Sec. \ref{sec:vct}. For instance, assume $U_i$ gives their vote to a candidate $B$ and therefore composes $\hat{v_i}=Enc_B (k_{B,i} || \texttt{IdB})$. 
The list of shuffled votes (obtained using the shuffling technique described in Sec. \ref {sec:shuffle}) is sent to the dealer.

\paragraph{P2SH address} The dealer, upon receiving the selected key shares from the $n$ voters, generates a P2SH address. The maximum number of public keys allowed in a multisignature output is 15 (see Eq.\ref{eq:script1}). Hence, 13 fields in the multisignature script can be used to store the vote of participants, while the two remaining fields can be used to store the public key given by a candidate ($M_A$ or $M_B$) as well as the public key assigned to the latter by the dealer ($P_A$ or $P_B$). 

The P2SH script is therefore based on $\it{if-else}$ $\it{statements}$, where each statement is a multisignature script incorporating up to 13 votes and the aforementioned keys of a candidate. A final statement, in the form of a $\it{scriptPubKey}$ ($S= k \times G$), is added in case no candidate obtains a sufficient number of votes after a period of time $\Delta T$. 
\paragraph{Voting Commitment Transaction} The dealer then creates the $\textit{Voting Commitment Transaction}$ (VCT) which includes one output where $n \times x$ Bitcoins are sent to the P2SH address. The transaction, illustrated in Fig. \ref{fig:tssv}, is sent to the voters who add their input (of $x$ Bitcoins) and sign the $\it{VCT}$. The last participant to sign the $\it{VCT}$ sends it to the dealer who will submit it for inclusion in the blockchain. Note that, while signing the $\it{VCT}$, the voters can verify in perfect anonymity the correctness of their vote included in the public record.  
\paragraph{Refund Transaction} Before the $\it{VCT}$ is actually broadcast on the network, the dealer creates a $\textit{Refund Transaction}$ (RT) that spends the output in the $\it{VCT}$ and redistributes $x$ Bitcoins to each voter. The transaction is signed by the dealer using the private key $k$ that unlocks the third option (although the latter includes a locking time preventing the UTXO to be spent before $\Delta T$). The dealer sends the signed $\it{RT}$ to the voters. Then and only then the $\it{VCT}$ is submitted for inclusion in the blockchain.

Ideally, the secret $k$ should be controlled by both the dealer and the voters so that under no circumstance the dealer can steal the voting fees. One could, for instance, add a $1$-of-$n$ multisignature in the script so that both the dealer's signature and at least one voter's signature is needed to spend the UTXO of the Voting Commitment Transaction.
\paragraph{Election result} In case one candidate successfully decrypts $t+1$ or more key shares in the shuffled list included in the P2SH script, the candidate can reconstruct the secret key needed to unlock the $n \times x$ Bitcoins in the $\it{VCT}$. If no candidate obtains sufficient key shares, the voters can broadcast the $\it{RT}$ and recover their fees.

\section{Discussion}\label{sec:end}
In the \textit{SHARVOT} protocol, transactions I/O shuffling and voting encryption techniques are implemented to guarantee the de-linking of the users from their vote submission, while conveniently constructed transactions ensure a permanent and immutable record of the ballot on the blockchain.

The dealer in $\it{SHARVOT}$ certifies the voters' eligibility through the distribution of key shares. The concatenation of the votes with an identifier for each candidate and the encryption of the vote prevents the dealer (or the voters) from manipulating the vote. Nevertheless, this does not completely prevent from the submission of multiple and/or erroneous votes by the same participant(s) with the malicious aim to disrupt the voting protocol. Such behaviour is however discouraged by attaching a cost, \textit{i.e.} the voting fee, to the voter's submission itself.
Dealer-less versions exist although these imply further rounds of communication among the voters and therefore would be more popular for an internal voting process with a limited number of participants.
Finally, a natural limitation of the protocol arises from the size of the script used to generate the P2SH output address in the $\it{VCT}$. 

\section{Conclusion}\label{sec:end2}
In this paper, we presented the \textit{SHARVOT} protocol that uses the blockchain technology to announce and build an election and determine a winning candidate by collecting voters' ballots in an immutable, storage-efficient and anonymous manner.

Indeed, we showed how the use of multisignatures in P2SH scripts allows for efficient storage of votes, which are encrypted using candidates identifiers and de-linked from the voters using shuffling techniques, and preserves the voters anonymity. Finally, the use of the blockchain guarantees data immutability and transparency. 

We also discussed the vulnerabilities associated to the SHARVOT protocol. Note that this solution remains secure, transparent and completely anonymous, without the need for complex cryptographic methods such as zero-knowledge proofs.


\bibliographystyle{ACM-Reference-Format}
\bibliography{sample-bibliography} 


\begin{thebibliography}{13}


\ifx \showCODEN    \undefined \def \showCODEN     #1{\unskip}     \fi
\ifx \showDOI      \undefined \def \showDOI       #1{#1}\fi
\ifx \showISBNx    \undefined \def \showISBNx     #1{\unskip}     \fi
\ifx \showISBNxiii \undefined \def \showISBNxiii  #1{\unskip}     \fi
\ifx \showISSN     \undefined \def \showISSN      #1{\unskip}     \fi
\ifx \showLCCN     \undefined \def \showLCCN      #1{\unskip}     \fi
\ifx \shownote     \undefined \def \shownote      #1{#1}          \fi
\ifx \showarticletitle \undefined \def \showarticletitle #1{#1}   \fi
\ifx \showURL      \undefined \def \showURL       {\relax}        \fi
\providecommand\bibfield[2]{#2}
\providecommand\bibinfo[2]{#2}
\providecommand\natexlab[1]{#1}
\providecommand\showeprint[2][]{arXiv:#2}

\bibitem[\protect\citeauthoryear{Antonopoulos}{Antonopoulos}{2014}]%
        {Antonopoulos}
\bibfield{author}{\bibinfo{person}{Andreas~M. Antonopoulos}.}
  \bibinfo{year}{2014}\natexlab{}.
\newblock \bibinfo{booktitle}{{\em Mastering Bitcoin: unlocking digital
  cryptocurrencies\/} (\bibinfo{edition}{1st.} ed.)}.
\newblock \bibinfo{publisher}{O'Reilly Media, Inc}.
\newblock


\bibitem[\protect\citeauthoryear{Nakamoto}{Nakamoto}{2008}]%
        {Satoshi}
\bibfield{author}{\bibinfo{person}{Satoshi Nakamoto}.}
  \bibinfo{year}{2008}\natexlab{}.
\newblock \showarticletitle{Bitcoin: A peer-to-peer electronic cash system.}
\newblock  (\bibinfo{year}{2008}).
\newblock


\bibitem[\protect\citeauthoryear{Riemann and Grumbach}{Riemann and
  Grumbach}{2017}]%
        {review1}
\bibfield{author}{\bibinfo{person}{Robert Riemann} {and}
  \bibinfo{person}{St\'ephane Grumbach}.} \bibinfo{year}{2017}\natexlab{}.
\newblock \showarticletitle{Distributed Protocols at the Rescue for Trustworthy
  Online Voting}.
\newblock \bibinfo{journal}{{\em ArXiv preprint\/}}
  \bibinfo{number}{1705.04480} (\bibinfo{year}{2017}).
\newblock


\bibitem[\protect\citeauthoryear{Chaum}{Chaum}{1981}]%
        {mixnet}
\bibfield{author}{\bibinfo{person}{David~L. Chaum}.}
  \bibinfo{year}{1981}\natexlab{}.
\newblock \showarticletitle{Untraceable electronic mail, return addresses, and
  digital pseudonyms}.
\newblock \bibinfo{journal}{{\it Commun. ACM}} \bibinfo{volume}{24},
  \bibinfo{number}{2} (\bibinfo{year}{1981}), \bibinfo{pages}{84--90}.
\newblock


\bibitem[\protect\citeauthoryear{Yao}{Yao}{2015}]%
        {MPC}
\bibfield{author}{\bibinfo{person}{Andrew~C. Yao}.}
  \bibinfo{year}{2015}\natexlab{}.
\newblock \showarticletitle{Protocols for secure computations.}
\newblock \bibinfo{journal}{{\em International Conference on Information and
  Communications Security,\/}} (\bibinfo{year}{2015}), \bibinfo{pages}{82--96}.
\newblock


\bibitem[\protect\citeauthoryear{Zhao and Chan.}{Zhao and Chan.}{2015}]%
        {ZKPvote}
\bibfield{author}{\bibinfo{person}{Zhichao Zhao} {and}
  \bibinfo{person}{T-H.~Hubert Chan.}} \bibinfo{year}{2015}\natexlab{}.
\newblock \showarticletitle{How to vote privately using bitcoin.}
\newblock \bibinfo{journal}{{\em International Conference on Information and
  Communications Security,\/}} (\bibinfo{year}{2015}),
  \bibinfo{pages}{160--164}.
\newblock


\bibitem[\protect\citeauthoryear{Katz and Lindell}{Katz and Lindell}{2014}]%
        {RSA}
\bibfield{author}{\bibinfo{person}{Jonathan Katz} {and} \bibinfo{person}{Yehuda
  Lindell}.} \bibinfo{year}{2014}\natexlab{}.
\newblock \bibinfo{booktitle}{{\em Introduction to modern cryptography}}.
\newblock \bibinfo{publisher}{CRC press}.
\newblock


\bibitem[\protect\citeauthoryear{Bartoletti and Pompianu}{Bartoletti and
  Pompianu}{2017}]%
        {OP-ret}
\bibfield{author}{\bibinfo{person}{Massimo Bartoletti} {and}
  \bibinfo{person}{Livio Pompianu}.} \bibinfo{year}{2017}\natexlab{}.
\newblock \showarticletitle{An analysis of Bitcoin OP\_RETURN metadata}.
\newblock \bibinfo{journal}{{\em ArXiv preprint\/}}
  \bibinfo{number}{1702.01024} (\bibinfo{year}{2017}).
\newblock


\bibitem[\protect\citeauthoryear{Shamir}{Shamir}{1979}]%
        {Shamir}
\bibfield{author}{\bibinfo{person}{Adi Shamir}.}
  \bibinfo{year}{1979}\natexlab{}.
\newblock \showarticletitle{How to share a secret}.
\newblock \bibinfo{journal}{{\it Commun. ACM}} \bibinfo{volume}{22},
  \bibinfo{number}{11} (\bibinfo{year}{1979}), \bibinfo{pages}{612--613}.
\newblock


\bibitem[\protect\citeauthoryear{Biryukov and Pustogarov}{Biryukov and
  Pustogarov}{2014}]%
        {attacks}
\bibfield{author}{\bibinfo{person}{Dmitry~Khovratovich Biryukov, Alex} {and}
  \bibinfo{person}{Ivan Pustogarov}.} \bibinfo{year}{2014}\natexlab{}.
\newblock \showarticletitle{Deanonymisation of clients in Bitcoin P2P network}.
\newblock \bibinfo{journal}{{\em Proceedings of the 2014 ACM SIGSAC Conference
  on Computer and Communications Security\/}} (\bibinfo{year}{2014}),
  \bibinfo{pages}{15--29}.
\newblock


\bibitem[\protect\citeauthoryear{ShenTu and Yu}{ShenTu and Yu}{2015}]%
        {mix}
\bibfield{author}{\bibinfo{person}{QingChun ShenTu} {and}
  \bibinfo{person}{JianPing Yu}.} \bibinfo{year}{2015}\natexlab{}.
\newblock \showarticletitle{Research on Anonymization and De-anonymization in
  the Bitcoin System}.
\newblock \bibinfo{journal}{{\em ArXiv preprint\/}}
  \bibinfo{number}{1510.07782} (\bibinfo{year}{2015}).
\newblock


\bibitem[\protect\citeauthoryear{Maxwell}{Maxwell}{2013}]%
        {coinj}
\bibfield{author}{\bibinfo{person}{Greg Maxwell}.}
  \bibinfo{year}{2013}\natexlab{}.
\newblock \showarticletitle{CoinJoin: Bitcoin privacy for the real world.}
\newblock \bibinfo{journal}{{\em Post on Bitcoin Forum\/}}
  (\bibinfo{year}{2013}).
\newblock
\showURL{%
Retrieved Feb 4, 2018 from
  \url{https://bitcointalk.org/index.php?topic=279249.0}}


\bibitem[\protect\citeauthoryear{Ruffing and Kate.}{Ruffing and Kate.}{2014}]%
        {coinsh}
\bibfield{author}{\bibinfo{person}{Pedro Moreno-Sanchez Ruffing, Tim} {and}
  \bibinfo{person}{Aniket Kate.}} \bibinfo{year}{2014}\natexlab{}.
\newblock \showarticletitle{CoinShuffle: Practical decentralized coin mixing
  for Bitcoin}.
\newblock \bibinfo{journal}{{\em European Symposium on Research in Computer
  Security\/}} (\bibinfo{year}{2014}), \bibinfo{pages}{345--364}.
\newblock


\end{thebibliography}

\end{document}